\documentclass[aps,pre,preprint,groupedaddress]{revtex4-1}

\begin{document}


\title{Viscous Dissipation in 2D Fluid Dynamics as a Symplectic Process and its Metriplectic Representation}


\author{Richard Blender}
\affiliation{Meteorological Institute, University of Hamburg, 
Hamburg, Germany}
\email[]{richard.blender@uni-hamburg.de}

\author{Gualtiero Badin}
\affiliation{Institute of Oceanography, University of Hamburg, Hamburg, Germany}
\email[]{gualtiero.badin@uni-hamburg.de}


\date{\today}

\begin{abstract}
Dissipation can be represented in Hamiltonian mechanics
in an extended phase space as a symplectic process.
The method uses an auxiliary variable
which represents the excitation of unresolved dynamics
and a Hamiltonian for the 
interaction between the resolved dynamics and the auxiliary variable. 
This method is applied to 
viscous dissipation (including hyper-viscosity) 
in a two-dimensional fluid, for which the dynamics is non-canonical. 
We derive a  metriplectic representation
and suggest a measure for the entropy of the system.
\end{abstract}


\maketitle


\section{Introduction}

Physical processes that conserve the total energy can be described 
in a Hamiltonian representation, 
in which the phase space volume is conserved. Hamiltonian dynamics 
is represented by a Poisson bracket which conserves a 
Hamiltonian function and satisfies Liouville's Theorem. 
The dynamics so obtained is based on symplectic geometry \cite{arnol2001symplectic}.
If the Poisson bracket is nonsingular, the Hamiltonian is a conserved quantity 
and the structure is also denoted as canonical. 
Otherwise, if the bracket is singular, the dynamics is called 
noncanonical and additional conservation laws exist, 
that take the name of Casimir functions.  
A classic example of singular Poisson brackets is given by 
fluid dynamics \cite{littlejohn82}, where there are infinitely 
many Casimir functions,
two notable examples of which 
are given by the enstrophy and the helicity. 

Lagrangian and Hamiltonian mechanics is traditionally
not able to represent dissipative processes like friction
or viscosity in terms of macroscopic variables.
Dissipative processes decrease kinetic energy and contract the volume in phase pace. 
There are two common approaches to embed dissipative processes 
in a geometric framework \cite{morrison09}: 
{\it gradient} systems and {\it metriplectic} systems. 
Gradient systems, also called incomplete systems, represent the dynamics 
by the gradient of a potential which is clearly not divergence-free. 
They dissipate energy and, possibly, other physical quantities and, 
by Lyapunov's theorem, have built-in asymptotic stability. 
A famous example of a gradient system is given by the Navier-Stokes 
equations. A special case of a gradient system applied to 
fluid dynamics 
and geophysical fluid dynamics is given by \cite{vallisetal89,shepherd90,GayBalmazHolmes13}, 
in which dissipation of energy and conservation of a Casimir 
(either helicity or enstrophy depending if the system is 2D or 3D) 
is used to find exact solutions of the Euler equations without 
changing the topology of the flow. 

Metriplectic systems, also called complete systems, 
are combinations of a symplectic and a metric bracket. 
Metriplectic systems are formulated by a noncanonical symplectic bracket
with an entropy-like Casimir \cite{morrison09}, and a metric bracket describing diffusion. Heuristically, the relationship between the Casimirs and the entropy in fluid dynamics is given by the fact that the Casimirs are associated to a relabeling symmetry \cite{padhye1996relabeling,Morrison-RMP-1998}, and hence to a counting of states \cite{morrison09}.
The symplectic Poisson bracket conserves the Hamiltonian and entropy 
while a metric bracket preserves the Hamiltonian and increases entropy. 
The conservation of energy and the increase of entropy are representation 
of the first and second laws of thermodynamics respectively. 
Metriplectic systems are typical in standard kinetic equations \cite{kaufman84,morrison1984bracket,morrison86,grmela1986bracket,turski1987canonical,beris1990poisson,holm2008kinetic}. 
For an example applied to geophysical fluid dynamics see \cite{Bihlo2008}.

A third way to include dissipation in Hamiltonian system is through the inclusion of stochastic noise. A complete list of works that employed this approach would be too long, and even sidetracking. A few notable examples will however be reported as a reference. A framework to include the statistical dynamics of a classical random variable that satisfies nonlinear equations of motion was introduced by \cite{martin1973statistical} and further developed by \cite{phythian1975operator,phythian1976further,phythian1977functional}, in which the author proposed a formulation of the closure problem based on generating functionals for the moments. The Langevin and Fokker-Plank equations were cast in a canonically invariant formulation by \cite{cepas1998canonically}. Through the inclusion of noise, a representation for dissipation as a gradient system was introduced by \cite{graham1984existence}, and then generalised by the same authors to include the cases of weak noise \cite{graham1984weak} and nondifferentiable potentials \cite{graham1985weak}. None of these studies were however extended to the case of fluid dynamics. 
Dissipation in fluids was introduced using the formalism proposed by \cite{martin1973statistical} e.g. by \cite{carnevale1983viscosity}. The use of field theory methodology for fluids includes also the study of the effective diffusivity for Gaussian turbulence in the near-Markovian limit, which was studied by \cite{drummond1982path} using path-integral methods. In a similar way, path-integral methods were used by e.g. \cite{carnevale1982field,navarra2013path} for geophysical flows. Closures for dissipation for fluid flows based on partial probability distributions were proposed e.g. by \cite{lundgren1972statistical}. 
Notice that while some of these frameworks were able to include stochastic noise in Hamiltonian systems, none of them put it in a symplectic form, which is one of the aims of this work. 

To 
represent dissipative processes like friction or viscosity in terms of macroscopic variables,
a method based on the extension of phase space 
was suggested in 
\cite{Morse_Feshbach1953}. 
The main ingredient proposed by \cite{Morse_Feshbach1953} is the introduction of an auxiliary variable
which grows in the time-reversed model with
negative friction or, in the case of fluid dynamics, negative viscosity. The method thus consists in finding the adjoint equation of the system \cite{atherton1975existence,nonnenmacher1986functional}. 
This auxiliary variable has no predefined physical meaning and in this study we will identify it with 
the excitation of unresolved (also called subgrid or subscale) processes. The same method was used, for example, in \cite{shah2015conservative} to study dissipation in canonical systems exhibiting limit cycles, and in \cite{Celeghini1992} for an application to quantum diffusion. For a review of other methods to represent frictional dissipation due to unresolved scales, see e.g. \cite{vanossi2013colloquium}. 
Here we apply the interaction Hamiltonian of \cite{Morse_Feshbach1953} 
for the dissipative interaction between the macrocopic variables 
and the auxiliary variable
in
2D fluid dynamics.  

To do so, we propose an interaction Hamiltonian which has form analogous to
the so-called $H=xp$ Hamiltonian, which yields unbounded hyperbolic trajectories in phase space and  
which has attracted particular interest, as it is conjectured to 
have a relationship with the zeros of the Riemann function 
 \cite{BerryKeating99,BerryKeating99b,Sierra11}.
Note that the interaction Hamiltonian
is unrelated to the conservative dynamics.
Reviews of different approaches to incorporate dissipation in Hamiltonian systems can be found in \cite{Riewe1996,sieniutycz2012conservation}.

\section{Two-dimensional fluid dynamics with viscous dissipation}
\label{sec:viscosity}


Two-dimensional incompressible fluid dynamics with viscous diffusion can be
described by the vorticity equation 
\begin{equation}
	\frac{ \partial \omega }{\partial t} 
	= J \left( \psi, \omega \right) + \nu \nabla^2 \omega~.
	\label{eq:31}
\end{equation}
The Jacobian  $J(a,b)=a_x b_y- a_y b_x$  is used to represent 
the advection of 
the vorticity  $\omega= \nabla^2 \psi$  by the flow, 
$(u,v)=(-\partial_y \psi,\partial_x \psi)$,
where  $\psi$ is the stream-function. 
In (\ref{eq:31}) $\nu$ is the (kinematic) viscosity. 
The vorticity equation can be described as a noncanonical 
Hamiltonian system
\cite{Morrison-RMP-1998}
\begin{equation}
	\frac{ \partial \omega }{\partial t} 
	= \{ \omega , \mathcal{H} \}_\mathcal{E}~,
	\label{eq:32}
\end{equation}
with the Hamiltonian in a periodic domain given by the kinetic energy 
\begin{equation}
	\mathcal{H} 
	= \frac{1}{2} \int \left( \nabla \psi \right)^2 dA ~
	= - \int \psi \omega dA~.
	\label{eq:33}
\end{equation}
with functional derivative $\mathcal{H}_{ \omega } = - \psi$, where the notation $\mathcal{H}_{\omega} = \delta \mathcal{H} / \delta \omega$ has been used.
The noncanonical Poisson bracket 
\begin{equation} \label{PoissonBracket}
	\{ \mathcal{F}, \mathcal{H} \}_\mathcal{E} = 
	\int \omega J\left( \mathcal{F}_{\omega},
	\mathcal{H}_{\omega} \right) dA~,
\end{equation}
has enstrophy $\mathcal{E}$ as a Casimir  
\begin{equation}
	\mathcal{E} = \frac{1}{2} \int \omega^2 dA ~,
	\label{eq:34}
\end{equation}
since $\left\{ \mathcal{F} [ \omega ],  \mathcal{E} \right\}_\mathcal{E} = 0$  
for any functional $\mathcal{F} [ \omega ]$.   
The enstrophy has functional derivative $\mathcal{E}_\omega  =  \omega$. 
This Casimir follows from the
particle relabeling symmetry \cite{padhye1996relabeling,Morrison-RMP-1998}.
In the following, the bracket (\ref{PoissonBracket}) will be called advective bracket.

\section{Interaction functional}

To rewrite the viscous diffusion of $\omega$, we introduce an 
interaction functional with an auxiliary variable $\sigma$
\begin{equation}
	\mathcal{M} = \int \omega \sigma dA~,
	\label{eq:38}
\end{equation}
with functional derivatives $\mathcal{M}_\omega =  \sigma$, $\mathcal{M}_\sigma  = \omega$. 
The variable $\sigma$ is an unspecified  
field representing local subscale processes.
The interaction functional (\ref{eq:38}) is written in analogy with the Hamiltonian interaction functional introduced by    
\cite{Morse_Feshbach1953} for the overdamped harmonic oscillator.
The trajectories induced by $\mathcal{M}$ are hyperbolic and unbounded, 
so that dissipation is represented as an irreversible process. 

\section{Anti-symmetric diffusive brackets}

We define an anti-symmetric bracket 
for two functionals $\mathcal{V}$ and $\mathcal{W}$ 
depending on $\omega$ and  $\sigma$ 
\begin{equation}
	\left\{ \mathcal{V}, \mathcal{W} \right\}' 
	=  \nu \int \left[\mathcal{V}_{\omega} \nabla^2 \mathcal{W}_{\sigma}  
	-  \mathcal{V}_{\sigma} \nabla^2 \mathcal{W}_{\omega} \right] dA ~.
	\label{eq:35}
\end{equation}	
The bracket (\ref{eq:35}) is anti-symmetric due to 
the symmetry of $\nabla^2$, 
\begin{equation}
	\left\{\mathcal{V}, \mathcal{W} \right\}' 
	= - \left\{ \mathcal{W}, \mathcal{V} \right\}'~,
	\label{eq:37}
\end{equation}
so that $\left\{\mathcal{V}, \mathcal{V} \right\}' = 0$. Since $\omega$ and $\sigma$ are independent degrees of freedom
we have 
\begin{equation} \label{omegasigma0}
	\{ \omega, \sigma \}' = 0~.
\end{equation}
It should be noted that, although (\ref{eq:35}) is used to describe dissipation,
it is not a metric bracket. 
The viscous diffusion term for the vorticity is obtained through 
this bracket $\left\{\cdot  , \cdot \right\}'$ and the functional $\mathcal{M}$,
so that
\begin{equation}
	\left.
	\frac{\partial \omega}{\partial t} 
	\right|_{visc}
	= \left\{ \omega , \mathcal{M} \right\}' = \nu \nabla^2 \omega~.
	\label{eq:39}
\end{equation}
To distinguish $\{ \cdot, \cdot\}'$ from the Poisson bracket (\ref{PoissonBracket}) 
we denote it as diffusive bracket.
A Poisson bracket  satisfies two conditions,
the Leibniz rule (or derivation property) 
and the Jacobi identity.
It can be proven that the diffusive bracket satisfies the Leibniz rule
which is similar to the product rule for differentiation,
\begin{equation} \label{Leibniz}
	\{\mathcal{F}\mathcal{G}, \mathcal{W} \}' 
	= 
	\mathcal{F} \{\mathcal{G}, \mathcal{W} \}' 
	+	
	\mathcal{G} \{\mathcal{F}, \mathcal{W} \}' ~,
\end{equation}
for arbitrary functionals $\mathcal{F}$, $\mathcal{G}$ and
$\mathcal{W}$.
The Jacobi identity, however, is not satsfied
\begin{equation} \label{Jacobi}
	\{\mathcal{F},  
	\{\mathcal{G}, \mathcal{W} \}'
	\}' 
	+
	\{\mathcal{G}, 
	\{\mathcal{W}, \mathcal{F} \}'
	\}' 
	+
	\{\mathcal{W}, 
	\{\mathcal{F}, \mathcal{G} \}'
	\}' 
	\neq 0~,
\end{equation}
as can be seen by simple substitution.
Therefore, the diffusive bracket is not a Poisson-bracket. 

The auxiliary variable  $\sigma$  grows like a 'negative viscosity'
\begin{equation}
	\frac{\partial \sigma}{\partial t} = \left\{ \sigma , \mathcal{M} \right\}' 
	= - \nu \nabla^2 \sigma~.
	\label{eq:310}
\end{equation}
In this framework the viscous decay of any functional  
$ \mathcal{F} \left[ \omega \right]$  
is determined by the bracket $ \left\{  \mathcal{F}, \mathcal{M} \right\}' $.
Equation (\ref{eq:310}) is the adjoint of (\ref{eq:39}).
The kinetic energy, given by the Hamiltonian $\mathcal{H}$, decays as
\begin{equation}
	\frac{d \mathcal{H}}{d t} 
	= \left\{ H , \mathcal{M} \right\}' 
	= - \nu \int 
	\mathcal{H}_{\omega} \nabla^2 \mathcal{M}_{\sigma} 
	dA~,
	\label{eq:311}
\end{equation}
since $\mathcal{H}$ does not depend on $\sigma$. 
This decay is proportional to the enstrophy 
\begin{equation}
	\frac{d \mathcal{H}}{d t} 
	=  \nu \int \psi \nabla^2 \omega dA  
	= - \nu \int \omega^2 dA ~
	= -2 \nu \mathcal{E}~.
	\label{eq:312}
\end{equation}
The enstrophy decays as
\begin{equation} \label{dEdtM}
	\left\{ \mathcal{E}, \mathcal{M} \right\}' 
	= \nu \int \omega \nabla^2 \omega dA \leq 0~,
\end{equation}
since the eigenvalues of $\nabla^2$ are negative in a periodic domain.

In the numerical modelling of geophysical flows, instead of viscosity, 
hyper-viscosity (or hyper-diffusion)
is frequently used which, compared to (\ref{eq:31}), 
is concentrated at high wave numbers with a reduced impact at low wave numbers 
\begin{equation}
	\left.
	\frac{\partial \omega}{\partial t} 	
	\right|_{hyp-visc}
	= \left( -1 \right)^{(n+1)} \nu \nabla^{2n} \omega~.
	\label{eq:313}
\end{equation}
Equation (\ref{eq:313}) includes Rayleigh friction  $- \nu \omega$  for $n=0$.
Hyper-viscosity is represented by a similar bracket as (\ref{eq:35}), 
with the same functional $\mathcal{M}$
\begin{equation}
	\left\{ \mathcal{F} , \mathcal{M} \right\}'' 
	= \left( -1 \right)^{(n+1)} 
	\nu \int \left[ \mathcal{F}_{\omega} \nabla^{2n} \mathcal{M}_{\sigma} 
	- \mathcal{F}_{\sigma} \nabla^{2n} \mathcal{M}_{\omega} \right] dA~.
	\label{eq:314}
\end{equation}

\section{Decomposition into symmetric brackets}

The bracket $\left\{\mathcal{F}, \mathcal{M} \right\}' $ can be separated 
into two terms by replacing $\mathcal{M}$ 
by two independent integrals, 
the enstrophy (\ref{eq:34}) and the quantity
\begin{equation}	 
	\mathcal{B} = \frac{1}{2} \int \sigma^2 dA~.
	\label{eq:317}
\end{equation}
Clearly $\mathcal{B}_\sigma=\sigma$ and $\mathcal{B}_\omega=0$. If we replace the functional derivatives of  $\mathcal{M}$  by those of 
$\mathcal{E}$ and $\mathcal{B}$,
\begin{equation}
	\mathcal{E}_{\omega} 
	= \mathcal{M}_{\sigma}~,
	~ \mathcal{B}_{\sigma} = \mathcal{M}_{\omega} ~,
	\label{eq:318}
\end{equation}
we obtain two symmetric brackets which involve derivatives of  $\omega$  and $\sigma$ only
\begin{eqnarray}
	\left\{ \mathcal{F} , \mathcal{M} \right\}' 	
	= \nu \int \left[ \mathcal{F}_{\omega} \nabla^2 \mathcal{E}_{\omega} 
	- \mathcal{F}_{\sigma} \nabla^2 \mathcal{B}_{\sigma} \right] dA \nonumber \\ 
	= \nu \int \mathcal{F}_{\omega} \nabla^2 \mathcal{E}_{\omega} dA 
	- \nu \int \mathcal{F}_{\sigma} \nabla^2 \mathcal{B}_{\sigma} dA \nonumber \\ 
	=  \left< \mathcal{F}, \mathcal{E} \right>_{\omega} 
	-  \left< \mathcal{F}, \mathcal{B} \right>_{\sigma}~.
	\label{eq:319}
\end{eqnarray}

The two functionals $\mathcal{E}$ and $\mathcal{B}$ 
split the dynamics in a decaying direction (for $\omega$)
and an expanding direction (for $\sigma$).
The geometric structure changes qualitatively 
since an anti-symmetric bracket is replaced by two symmetric brackets.
This decomposition is comparable to the decomposition of a 
Nambu bracket  by so-called constitutive conservation laws
\cite{NevirSommer2009, SalazarKurgansky2010, BlenderBadin15}.
The $\omega$-bracket yields diffusion, 
\begin{equation}
	\left.
	\frac{\partial \omega}{\partial t} 
	\right|_{visc}
	= \left< \omega , \mathcal{E} \right>_{\omega} 
	= \nu \nabla^2 \omega ~,
	\label{eq:320}
\end{equation}
and the $\sigma$-bracket yields the growth of the auxiliary variable $\sigma$
\begin{equation}
	\frac{\partial \sigma}{\partial t} 
	= - \left< \sigma , \mathcal{B} \right>_{\sigma} 
	= - \nu \nabla^2 \sigma ~.
	\label{eq:321}
\end{equation} 
The difference $\Phi = \mathcal{B} - \mathcal{E}$  
grows according to
\begin{eqnarray}
	\frac{d \Phi}{d t} 
	= \left\{ \Phi, \mathcal{M} \right\}' 
	= \left\{ \mathcal{B},  \mathcal{M} \right\}' 
	-  \left\{ \mathcal{E},  \mathcal{M} \right\}' \nonumber \\
	= - \left< \mathcal{B},  \mathcal{B} \right>_\sigma 
	-  \left< \mathcal{E},  \mathcal{E} \right>_\omega \geq 0~.
	\label{eq:322}
\end{eqnarray} 

To summarize, the main properties of the two brackets are:
(i) The advective bracket preserves all integrals
\begin{equation}
	\{ \mathcal{H}, \mathcal{H} \}_\mathcal{E} = 0~, ~
	\{ \Phi,        \mathcal{H} \}_\mathcal{E} = 0~,
	\label{eq:323}
\end{equation} 
with the conservation laws
\begin{equation}
\{ \mathcal{E}, \mathcal{H} \}_\mathcal{E} = 0~, ~
\{ \mathcal{B}, \mathcal{H} \}_\mathcal{E} = 0. 
	\label{eq:323a}
\end{equation} 
And (ii), the diffusive bracket impacts all integrals
\begin{equation}
	\{ \mathcal{H}, \mathcal{M} \}' \leq 0~, ~
	\{ \Phi,        \mathcal{M} \}' \geq 0~ ,
	\label{eq:324}
\end{equation} 
and 
\begin{equation}
\{ \mathcal{E}, \mathcal{M} \}'   \leq 0~, ~
\{ \mathcal{B}, \mathcal{M} \}'   \geq 0~.
	\label{eq:324a}
\end{equation}

\section{Metriplectic representation of dynamics}

The dynamics of an arbitrary functional $\mathcal{F}$ of 
vorticity $\omega$ and the auxiliary variable $\sigma$ is given by
the non-canonical Poisson bracket (\ref{PoissonBracket})
and the diffusive bracket
\begin{equation} \label{dFdt}
	\frac{\partial}{\partial t} \mathcal{F}
	= \{\mathcal{F},\mathcal{H} \}_\mathcal{E} +  \{ \mathcal{F}, \mathcal{M} \}'~.
\end{equation}

Using this representation for the dynamics, it is possible to derive a metriplectic formulation 
based on (\ref{dFdt}) if $\mathcal{F}$ does not depend on $\sigma$ 
and the dynamics is restricted to the vorticity $\omega$ 
\begin{equation} \label{dFwtdt}
	\frac{\partial}{\partial t} \mathcal{F}
	= \{\mathcal{F},\mathcal{H} \}_\mathcal{E} 
	+  
	\left< \mathcal{F}, \mathcal{E} \right>_{\omega} ~.
\end{equation}
This representation uses a symplectic Poisson-bracket and a 
metric bracket defined in (\ref{eq:319}).
The role of the enstrophy is thus two-fold: it is a Casimir of the Poisson-bracket  
and determines the dissipation in the metric bracket.	
This is the formulation of metriplectic systems suggested by \cite{kaufman84}.
However, the Hamiltonian (the kinetic energy) is not preserved in the diffusive bracket. 
This is similar to the decay of the Hamiltonian found by \cite{Bihlo2008} 
in the metriplectic form of Rayleigh-B\'enard convection with viscosity 
and temperature diffusion.

In metriplectic systems the Casimir functions can be 
considered as candidates for entropy  \cite{morrison09}, due to the fact that the Casimirs are associated to a relabeling symmetry and hence to a counting of states.
The metriplectic representation (\ref{dFwtdt}) suggests to
consider the (negative) enstrophy as entropy.

\section{Summary and Discussion}
\label{sec:discussion}

In this paper we reconsider the description of dissipative processes 
 based on an extended phase space
\cite{Morse_Feshbach1953}.
An auxiliary variable is added which represents sub scale degrees of freedom
and evolves according to time-reversed equations.
Further, an interaction Hamiltonian yielding unbounded hyperbolic trajectories 
in the canonical phase space, 
which we consider as a necessary property of irreversible dissipative processes, is considered. 
The dynamics of the extended system 
is nondivergent in phase space and satisfies Liouville's Theorem. 
The resulting formulation of dynamics yields thus
a symplectic description of dissipation. 
As an application of this approach we have considered an infinite dimensional system, where dissipation takes the form of viscosity in two-dimensional fluid dynamics 
(with an extension to hyperviscosity or hyperdiffusion). 
From a geometric point of view, the system considered here is neither an
incomplete gradient nor a complete metriplectic system. 

The symplectic description of viscosity uses an interaction Hamiltonian
given by the bilinear coupling of vorticity and the auxiliary variable.
An anti-symmetric bracket (which is not a Poisson bracket)
reveals diffusion for the vorticity and 'anti-diffusion' (e.g. a negative viscosity) for the auxiliary variable.
By a redefinition of the interaction Hamiltonian
this bracket can be split into two symmetric brackets, giving rise to a metriplectic system.
In this case, enstrophy appears as a Casimir of the non-canonical bracket
and in the symmetric bracket which is responsible for dissipation.
This suggests to consider the negative enstrophy as a measure for entropy.

As an application, the formulation that was presented in this work, would allow for a consistent representation of dissipation in numerical discretizations based on symplectic integrators \cite{leimkuhler2004simulating}. This is particularly interesting in the case of 2D and geostrophic turbulence:  \cite{sommer2009conservative} showed that a discretization based on the Nambu brackets yields different slopes for the energy spectra due to the elimination of spurious sources of enstrophy. Because turbulence is essentially a dissipative process, the inclusion of a consistent symplectic scheme for the dissipation might have an important impact on the dynamics. 
In this regard, relations (\ref{eq:323})-(\ref{eq:324a}) could be used to evaluate the behaviour of the discretization. This next step will be part of future work.

\begin{acknowledgments}
The authors would like to thank and anonymous referee for constructive comments.
This research partially funded by the research grant DFG TRR181. GB was partially funded also by the research grant DFG 1740.
\end{acknowledgments}


\providecommand{\noopsort}[1]{}\providecommand{\singleletter}[1]{#1}%

\end{document}